\documentstyle[12pt]{article}
\renewcommand{\baselinestretch}{1.5}
\input epsf

\let\a=\alpha \def\b{\bar{\a}} \let\g=\gamma \let\d=\delta
\let\e=\epsilon   
  \def\k{\vec{k}} \let\l=\lambda
\let\m=\mu \let\n=\nu \let\x=\xi \def\p{\vec{p}} \def\q{\vec{q}}
\let\r=\rho \let\s=\sigma \let\t=\tau \let\o=\omega 
   
 \let\S=\Sigma   \let\L=\Lambda
\let\G=\Gamma \let\D=\Delta  

\def\2{{1\over2}} \def\4{{1\over4}} \def\52{{5\over2}} \def\6{\partial }
\def\({\left(} \def\){\right)} \def\<{\langle } \def\>{\rangle }

\def\CL{{\cal L}}
\def\CP{{\cal P}}

\def\beg{\begin{equation}}
\def\begar{\begin{eqnarray}}
\def\ee{\end{equation}}
\def\ea{\end{eqnarray}}

\newcommand{\pref}[1]{(\ref{#1})}                
\newcommand{\plabel}[1]{\label{#1}}              
\newcommand{\pcite}[1]{\cite{#1}}                
\newcommand{\pbib}[1]{\bibitem{#1}}              
\renewcommand{\Im}{\mbox{Im}}
\newcommand{\ps}{p\!\!\!/}                       
\newcommand{\pss}{p\!\!\!/\,'}                   

\begin{document}

\begin{titlepage}

\topmargin-2cm

\hfill TUW-95-02
\vspace{.7cm}
\begin{center}
\Large
{\bf \renewcommand{\baselinestretch}{.5} Relativistic Bound State
Equation for Unstable Fermions and the Toponium Width} \\[1cm]
\normalsize
Wolfgang Kummer and Wolfgang M\"odritsch$^1$ \\[.6cm]
{\renewcommand{\baselinestretch}{1} \small Institut f\"ur Theoretische Physik\\
Technische Universit\"at Wien\\ Wiedner Hauptstra\ss e 8-10/136, A-1040 Wien\\
Austria\\}
\end{center}
\vspace{1cm}

\centerline{{\bf Abstract}}

\vspace{.7cm}
\parbox{16cm}
{\renewcommand{\baselinestretch}{1} The bound state problem
for a fermion-antifermion system is considered taking into account
a finite decay width of the constituents. We propose an exactly
solvable relativistic zero order equation similar to that of Barbieri and
Remiddi, but including a constant width.  We focus especially on the $t\bar{t}$
system for which we reconsider our recent calculation of the bound state
corrections to the toponium width, which was performed in the narrow width
approximation and needed the use of second order Bethe-Salpeter perturbation
theory.  We show that one obtains the same result already in first order BS
perturbation theory if one uses our present approach. Furthermore the large
cancellations of gauge dependent terms is demonstrated to be a consequence of a
Ward identity. This cancellation mechanism is shown to be valid for
general fermion-antifermion systems. \baselineskip22pt }

\vspace{1cm}

\tiny
\_\hrulefill \hspace*{8cm} \\
Vienna, January 1995\\
$^1$e-mail: wmoedrit@ecxph.tuwien.ac.at, FAX: +43-1-5867760

\end{titlepage}

\newpage
\renewcommand{\baselinestretch}{2}
\normalsize

1. The possible importance of off-shell corrections to the decay width in the
calculation of the cross section for top-antitop production by $e^+ e^-$
lately has been the subject of detailed analysis \pcite{Sumino,Jeza}.
In a recent publication \pcite{topdec} we addressed this problem in the
context of
Bethe-Salpeter(BS) perturbation theory starting from the Barbieri-Remiddi
equation \pcite{BR}.  We were able to show by explicit calculation that large
cancellations took place and that the net result can be interpreted as a pure
time dilatation effect.  This had been conjectured qualitatively before
\pcite{Jeza} in view of the known
results from muonium \pcite{Ueber}.

In this note we will improve our understanding of this result and derive a
general theorem, valid for similar cases.  From a theoretical point of view our
calculation for toponium has been unsatisfactory since it was obtained in
the narrow width approximation. As we will explain below this approximation
is not strictly applicable to that system.
We now include the "bulk" of the decay width already in the
zero order equation to avoid this problem. This offers also the more technical
advantage that we then need only to consider first order BS- perturbation
theory
in contrast to the calculation in the narrow width approximation where second
order perturbation theory was needed.  Furthermore we are able to show that
the cancellation of the gauge dependent terms is a result of a Ward identity.
This fact also considerably simplifies the actual calculation.


2. The Bethe-Salpeter (BS) approach for weakly bound systems is based upon the
BS-equation
\beg
G=D+DKG \plabel{BS}
\ee
for the $t\bar{t}$ Green function $G(P,k,k')$, where $D$ is the product of the
two (full) propagators and $K$ represents the 2pi BS-kernel. All four point
functions in \pref{BS} depend on the total momentum $P = (P_0,\vec{0})$; the
incoming (outgoing) lines carry momentum $P/2 \pm k$ ($P/2 \pm k'$).

Perturbation theory starts from an equation similar to eq. \pref{BS} with
$G_0,D_0$ and $K_0$ chosen in such a way that the exact solution is known. If
$D_0$ consists of the nonrelativistic fermion-propagators only and if $K_0$ is
the Coulomb kernel the zero order equation corresponding to \pref{BS} (after
integrating out $k_0$ and $k'_0$) simply reduces to the Schr\"odinger equation.
Another very convenient zero order equation which already includes the
relativistic free fermion propagators and still remains solvable has been found
by Barbieri and Remiddi (BR) some time ago \pcite{BR}.  However,
as is the case for the top quark - if the decay
width of the constituents becomes comparable to the binding energy,
perturbation theory runs into troubles because of the occurrence of terms
$(\G/(\a^2 m))^n$ in graphs like the one in fig. 1.  To circumvent these
difficulties one has to include at least a part of the exact self energy
function in the zero order equation \pcite{toppot}.

The key point of the BS approach as generalized to the case of a particle with
finite width is the approximation to the free fermion propagator
$E_k = \sqrt{\vec{k}\,^2+m^2}$
\begar \plabel{Sg}
S_{\pm}(\pm \frac{P_0}{2}+k_0) &=& [(\pm \frac{P_0}{2}+k)\g - m - \S(\pm
\frac{P_0}{2}+k) ]^{-1} \to  \\
&\to& \frac{\L^+ \g_0}{\pm \frac{P_0}{2}+k_0-E_k+i \frac{\G}{2}} + \frac{\L^-
\g_0}{\pm \frac{P_0}{2}+k_0+E_k-i\frac{\G}{2} }, \plabel{D0}
\ea
which is surely valid for a slowly moving particle. This choice leaves
the relativistic projectors
\beg
\L^{\pm}(\p) \equiv \frac{E_p \pm (\vec{\a} \p + \beta m)}{2E_p}.  \plabel{L}
\ee
unaffected and thus introduces a minimal number of terms involving
the width $\G$ in the calculation.  By integrating the zero
order BS-equation for the Feynman amplitudes ( $D_0 = S_+ \otimes S_-$)
\beg \plabel{BSF}
i G_0 = - D_0 + D_0 K_0 G_0
\ee
with respect to
the zero components of momentum under the assumption that $K_0$ is
instantaneous and with a kernel of the special form ($ \b = 4/3 a_s$)
\beg  \plabel{K0}
K_{0} = [\g_0 \L^+ \l^+ \L'^+] \otimes [\L^- \l^- \L'^- \g_0 ] m \m \n \m' \n'
\frac{- 4\pi \b}{\vec{q}\,^2}
\ee
with
\begar
\m &=& 2E_k/(E_k+m) \nonumber \\ \n &=& 2/\sqrt{P_0+2E_k+i\G} \nonumber \\
\l^{\pm} &=& \2 ( 1 \pm \g_0 ) \nonumber
\ea
so as to annihilate the second term in \pref{D0} we can show that
\begar
G_0 &=& i (2\pi)^4 \d (k-k') D_0 + [\L^+ \l^+ \L'^+ \g_0 ] \otimes [ \g_0 \L^-
\l^- \L'^-] \times \\
& & \times \frac{2 \o}{k_0^2-\o^2} \m \m' \left[ \frac{(2 \pi)^3
\d(\k-\k\,')}{2 \o} -
\frac{G_C(\widehat{E},\k,\k\,')}{m \n \n'} \right] \frac{2 \o'}{k_0'^2-\o'^2}
\nonumber
\ea
with
\begar
\o &:=& E_k - \frac{P_0+i \G}{2}
\ea
is the solution of eq. \pref{BSF}. Primes denote the dependence on the
corresponding momenta and $G_C(\widehat{E},\k,\k\,')$ is the nonrelativistic
Coulomb Green function \pcite{Schw} evaluated at $$ \widehat{E} = \frac{1}{4m}[
(P_0+i \G)^2 - 4m^2 ].  $$ This solution shows that the width of a particle can
also consistently be included for a {\it relativistic} zero order equation
similar to the nonrelativistic case where the replacement $E \to E+i\G$ in the
Green function has been proposed first \pcite{Fadin}.

The corresponding solution for the BS-wave functions \pcite{toppot} $(\o_n =
E_p -M_n^{(0)}/2-i\e)$
\beg
\chi_n (p,\e)= \g_0 \bar{\chi}_n^*(p,-\e) \g_0 = i \frac{\L^+ S \g_0 \L^- \g_0}
{(p_0^2 - \o_n^2) } \frac{\m(p)}{\n(p)} \frac{2 \o_n}{
\scriptstyle{\sqrt{M_n^{(0)}}} }
\phi(\p). \plabel{xBR}
\ee
is identical to the BR wave-function for stable quarks and belongs to the
spectrum of bound states $ P_n^{(0)} =M_n^{(0)} - i \G_t = \sqrt{1-\s_n^2} -i
\G $. It appears as the
real residue of the complex pole.  The small imaginary quantity
 $i \e$ determines the integration around that pole.

In eqs. \pref{xBR} $S$ is a constant $4\times4$ matrix which represents the
spin state of the particle-antiparticle system:
\beg \plabel{spin}
S = \left\{ \begin{array}{l@{\quad:\quad}l} \g_5 \l^- & \mbox{singlet} \\
\vec{a}_m \vec{\g} \l^- &  \mbox{triplet}.
\end{array} \right.
\ee
$\phi$ is simply the normalized solution of the Schr\"odinger equation in
momentum space, depending on the usual quantum numbers $(n,l,m)$; $a_{\pm 1},
a_0$ in \pref{spin} describe the triplet states.  In the following it will
sometimes be sufficient to use the nonrelativistic approximations of eqs.
\pref{xBR} ($\o_n \approx \tilde{\o}_n = (\p\,^2/m+E_n)/2 -i\e $)
\begar
\chi(p,\e)^{nr} &=& \frac{\sqrt{2} i \tilde{\o}_n}{p_0^2-\tilde{\o}_n^2}
\phi(\p) S = \g_0 \bar{\chi}^*(p,-\e)^{nr} \g_0 \plabel{chinr}
\ea

Near a bound state, whose pole now lies in the lower half plane at
$P_0 = P_{n}^{(0)}$, the
Green function of the unperturbed (BR) equation behaves as
\beg  \plabel{Gp}
G_{0}(P,k,k') \approx \frac{ \chi(k) \otimes \bar{\chi}(k') } { P_0 -
P_{n}^{(0)}}
\ee
A generic perturbation $H$ changes $G_0$ into
\beg   \plabel{Gsum}
G=\sum_{\n=0}^{\infty} G_0 (H G_0)^{\n}.
\ee
Since $G,G_0,$ etc. denote Feynman amplitudes, $H$ in \pref{Gsum} differs
from such an amplitude by a minus sign.  It is straightforward to show that
\pcite{toppot}.
\beg \plabel{H}
H=-K+K_0+iD^{-1}-iD_0^{-1}
\ee
The l.h.s of eq. \pref{Gsum} looks like \pref{Gp} with
$P_{n}^{(0)}$ replaced by $P_n$ and with perturbed wave functions.

Formally expanding both sides of eq. \pref{Gsum} near a certain pole
$P_{n}^{(0)}$
\begar
G_0 &=& \sum_{m=0}^{\infty} g_m (P_0 - P_{n}^{(0)})^{m-1} \plabel{gent}\\ H &=&
\sum_{m=0}^{\infty} h_m (P_0 - P_{n}^{(0)})^{m} \nonumber
\ea
and using $g_0 = \chi \otimes \bar{\chi}$ allows the determination of the shift
of the position of the pole $\D P_n = P_n - P_n^{(0)}$ to arbitrary order
\beg
\D P_n = \D M_n - i \frac{\D \G_n}{2}=  \< \< h_0 \> \>
+\< \< h_0 g_1 h_0 \> \> + \< \< h_0 \> \> \< \< h_1 \> \> + O(h^3).
\plabel{pts}
\ee
A shorthand for BS expectation values has been used which incorporates
four-integrations of the relative momenta ( $\int dk = (2\pi)^{-4} \int d^4k$):
\begar
\< \< a \> \> &=& \int dk dk' \bar{\chi}_{ij}(k) a_{ii'jj'}(k,k')
\chi_{i'j'}(k'), \plabel{erww}
\ea
As already in eq. \pref{BS} etc., in composite expressions integrations and
summations over internal variables are understood. $i,j$ etc. represent spinor
as well as color indices. A similar formula as eq. \pref{pts} holds for the
perturbed wave functions \pcite{Lepage}. It is important to note that
eq. \pref{pts} is not a power series in $\a_s$.

3. The correction to the decay width is obtained from the imaginary part of the
expectation values, receiving contributions from the graphs in fig.2 and fig.3.
In ref. \pcite{topdec} the problem of the gauge independent calculation of
bound state corrections to the toponium decay width had been addressed. It was
shown
that it is possible to obtain a gauge independent result which can be
interpreted as a correction due to time dilatation. But since the top quark
decay width is of the same order of magnitude as the binding energy
it is obviously preferable to
use the equations valid for decaying particles as described above.
In our previous approach it was essential that $O(\a_s^2)$ corrections
arose from all three terms
denoted in \pref{pts} because the BR equation was used to zero order. Here we
will show that within our present formalism we need only to consider first
order BS perturbation theory e.g. the first term in \pref{pts}.

Corrections to the fermion propagator have to be included via the perturbation
kernel (see fig. 2)
\beg
h_0^{(1)} = ( i D^{-1} -i D_0^{-1} ) \Big|_{P_0 = M_n^{(0)} - i \G}.
\ee
We emphasize that the perturbation kernel has to be evaluated at
$P_0 = M_n^{(0)} - i \G$.
With equ. \pref{H} and $D^{-1}_0 \big|_{P_0 = M_n^{(0)} - i \G} = D^{-1}_0
\big|_{\G=0}$  we find
\beg \plabel{h1}
h_0^{(1)} = - i [ (\Sigma (p-\frac{i \G}{2} n)+i \frac{\G}{2} \g_0 ) \otimes
     (\pss-m) + (\ps-m)
   \otimes  (\Sigma (p'+\frac{i \G}{2} n)-i \frac{\G}{2} \g_0 ) ] (2\pi)^4
   \d(k-k')
\ee
where $n=(1,\vec{0})$ and $p=\frac{M_n^{(0)}}{2}+k$,
$-p'=\frac{M_n^{(0)}}{2}-k'$ denote the four momentum
of the quark and the antiquark, respectively. The direct product refers to the
$t \otimes \bar{t}$ spinor space. The factors $(\ps-m)$ and $(\pss-m)$
compensate the superfluous propagator on the line without $\S$. We have
neglected terms that will contribute to $O(\a^2 \G^2 /m)$.

Since we are calculating higher order effects we may also drop the dependence
on the mass of the bottom quark although it may be included in principle.  For
the electroweak theory we use the $R_{\xi}$-gauge with the gauge fixing
Lagrangian ( $M$ denotes the mass of the W-Boson) $$ \CL_{gf} = - \xi |\6^{\m}
W_{\m}^+ - i \frac{M}{\xi} \phi^+ |^2 $$ because it eliminates mixed $W-\phi$
propagators. The gauge parameter $\xi$ will not be fixed in the following.  We
obtain for the imaginary parts of $\S$ in \pref{h1} ($s=\sin \theta_W$,
$\CP_{\pm} = (1 \pm \g_5)/2 )$ for $p^2 > M^2$ and $p^2 \xi > M^2 $:
\begar
- \Im \S_W &=& \frac{e^2}{s^2} \frac{\ps}{16 \pi} [\CP_+ A^{(W)} + \CP_-
A^{(\phi)} ] \plabel{SW}
\ea
where
\begar
A^{(W)} &=& \t(p^2) + \r(p^2,\xi) \nonumber \\ A^{(\phi)} &=& m^2 (
\frac{\t(p^2)}{2 M^2} - \frac{\r(p^2,\xi)}{p^2} )
\plabel{As} \\
\t(p^2) &=&  \frac{(p^2-M^2)^2}{2 p^4}, \quad \r(p^2,\xi)=
\frac{1-\x}{2\x} ( 1- \frac{M^2}{2p^2}\frac{1+\x}{\x}) \nonumber
\ea

Now we observe that the terms in \pref{h1} which include $\S$ have already been
calculated in \pcite{topdec} except for the imaginary parts occurring in the
trace
\begar
T &:=& \2 \mbox{tr}[\g_0 \L^- S^{\dagger} \L^+\g_0 \CP_{\pm} \ps \L^+S \L^-
(\ps-m) ] = \nonumber \\ &\approx& \2 (p_0+\o)[
m(1-\frac{\p\,^2}{m^2})+(p_0-\o)-i \frac{\G}{2}].  \plabel{S}
\ea
and in
\begar
p^2-m^2 \approx 2m p_0 -\p\,^2 - m \s^2 -i m \G. \plabel{pmm}
\ea
We will be able to neglect these imaginary parts in the following since they
either give rise to real corrections ( which we are not interested in) or
to imaginary ones from Re$\S$, both of which are of $O(\a_{weak}^2)$.
Of course, these corrections would be required a in
complete calculation of the toponium decay width to this order in order to
obtain a gauge independent result.  This goes beyond the scope of the present
paper. Furthermore, these corrections are state independent, expected to be
very small and therefore of minor practical interest, anyhow.
As can be seen from \pref{S} the axial part of $\S$ does not contribute within
the expectation value.

Therefore, the contribution from first order BS perturbation theory to the
decay
width to $O(\a_s^2 \G)$ reads
\beg
\< \< h_0^{(1)} \> \> = \G_1 + \G_3
\ee
with
\begar
\G_1 &=& -4 \Im \< \< - i \Sigma (p) \otimes (\pss-m) (2\pi)^4 \d(k-k') \>\> =
\\
&=& \G_0 - 2 \G_0 \s_n^2 + \frac{e^2m}{16 \pi s^2} [( 2 M^2 + m^2 )
\frac{m^2-M^2}{m^4} + 2 \r(m^2,\xi)] \< \frac{p^2 -m^2}{m^2} \>,\nonumber  \\
\G_3 &=& -4 \Im \< \<  \frac{\G}{2}  \otimes (\pss-m) (2\pi)^4 \d(k-k') \>\> =
- \G_0 ( 1+ \frac{\s_n^2}{2})
\ea
A factor 2 arises here and in the following, counting both self energy
contributions of the $t$ and $\bar{t}$ line:
\beg \plabel{Gam0}
\G_0 = - 2 \Im \S^{(W)}(m) = \frac{e^2m}{16 \pi s^2} (1+\frac{m^2}{2M^2})
\frac{(m^2-M^2)^2}{m^4}
\ee
In the sum $\G_1 +\G_3$ the leading $\G_0$ terms cancel and the resulting
correction is already of $O(\a_S^2 \G_0)$.

Now we turn to the calculation of the vertex correction (Fig. 3).  Performing
the color trace this gives rise to the kernel
\beg \plabel{h2}
h_0^{(2)} = i \Im \L_0 \otimes \g_0 \frac{4\pi \b}{\vec{q}\,^2}
\ee

In \pcite{topdec} we calculated $\Im \L_0 := \lim_{q \to 0, p^2 \to m^2}
\L(p,q)$ directly. Here we will show
that it is possible to use a Ward identity to determine this correction.  We
observe that
\begar
\L_0^a(p,q) \Big|_{q \to 0 } &=& - g_{QCD} T^a \frac{\6}{\6 p_0} \S (p)
\plabel{WI}
\ea
holds at least at the one-loop level. In the graphs shown in fig. 3 to the
required
order it proves sufficient to evaluate the leading contribution by putting the
fermions on the mass shell and applying the nonrelativistic wave functions
\pref{chinr}.
Using the explicit expression \pref{SW}  for $\Im \S$ we obtain
\begar
\Im \L_0 &=& \frac{e^2}{s^2 16 \pi} \g_0 [ A^{(W)}(m^2) + A^{(\phi)}(m^2) +
2 m^2 \frac{\6}{\6 p^2} ( A^{(W)}(p^2) + A^{(\phi)}(p^2) ) \Big|_{p^2=m^2} ]
\nonumber \\
&=& \frac{e^2}{32 \pi s^2 } \g_0 F \\ F &=& (1+\frac{m^2}{2M^2}) (1-
\frac{M^2}{m^2})(1+ 3 \frac{M^2}{m^2}) + 4 \r(m^2,\xi).
\nonumber
\ea
This result agrees with that of ref. \pcite{topdec} and one obtains a
correction to the decay width:
\begar \plabel{Gam2}
\G_2 = \frac{e^2}{16 \pi s^2} \< \frac{4\pi\b}{\q^2} \> F.
\ea

The fact that the gauge dependent terms in the sum $\G _1 + \G_2$ cancel is
thus traced back to the identity \pref{WI}. The gauge independent contribution
from first order BS-perturbation theory is now different from that of
\pcite{topdec} (actually it gives now the net
result) in accordance with the different zero order equation, used in our
present context.

To complete our calculation we finally check the contributions of second order
perturbation theory. It can be shown that due to the fact that the first weak
correction now is already of order $O(\a_s^2 \G)$, this corrections do not
contribute to the required order.  Thus our present approach simplifies the
calculation of bound state corrections to the decay width considerably.


Summing up all contributions we get the result
\beg \plabel{result}
\D \G_{boundstate} = \G_1 +\G_2 +\G_3 = - \G_0  \frac{\s_n^2}{2}
\ee
in agreement with our previous calculation \pcite{topdec}.

4. The above considerations suggest that the result \pref{result} is only a
special case of a more fundamental statement, with broader range of
applicability, which we will now derive.

Consider a fermion whose decay can be described by the imaginary part of a self
energy function which will have the general form (in any covariant gauge in the
relevant sector of the theory)
\beg
\S(p) = \S_S(p^2) + \ps \S_V(p^2) + \g_5 \S_P(p^2) + \ps \g_5 \S_A(p^2).
\ee
Within the expectation value \pref{erww} the pseudoscalar and axial vector
parts vanish. Furthermore we will drop all factors $\Theta(p^2-\m^2)$, which is
a valid approximation in all cases where $\a^2 m \ll m- \m$. If this is not the
case, either the whole calculation makes no sense because then one would have
to
include the entire rung of Coulomb interactions of the remaining particle with
the heavy decay products or the decay width becomes very small, or both.
Let us therefore consider the correction to the decay width which results from
the remaining parts of the self energy in an on shell renormalization scheme (
$m$ is the pole mass of the fermion). We expand $\S$ around that mass shell:
\beg \plabel{sigall}
\S(p) = i (\Im \S_S + \ps \Im \S_V) + (\S_S'+\ps \S_V')(p^2-m^2) .
\ee
Terms of $O(\a_{weak}^2)$ are understood to be neglected (cf. remarks after
eq. \pref{pmm} ) and we denote
\begar
\S_X &:=& \S_X(m^2), \qquad X = S,V \nonumber \\ \S_X' &:=& \frac{\6}{\6 p^2}
\S_X(p^2) \Big|_{p^2=m^2}. \nonumber
\ea
{}From eqs. \pref{WI} and \pref{sigall} we can easily calculate the vertex
correction, say, for the upper particle line:
\beg
\L_0 \l^+ = g T^a \g_0 [i \Im \S_V + 2m (\S_S'+m \S_V')] \l^+
\ee
After some algebra we can write to the required order
\begar
h_0^{(\S)} &=& (\Im \S_S + m \Im \S_V + \frac{\G}{2}\g_0 ) \otimes (\pss-m)
- i D_{0,\G=0}^{-1} [i \Im \S_V + 2m (\S_S'+m \S_V')] \nonumber \\
\\
h_0^{(\L)} &=& -\frac{4 \pi \b}{\q\,^2} [i \Im \S_V + 2m (\S_S'+m \S_V')] \g_0
\otimes \g_0 \nonumber
\ea
Since
\beg
\G = -2 \bar{u} \Im \S u \Big|_{p = (m,\vec{0})}= - \Im \S_S(m^2) - m \Im
\S_V(m^2)
\ee
and $ (i D_{0,\G=0}^{-1} - K_{0,\G=0}) \chi = 0$ we obtain for the bound state
correction to the decay with to $O(\a_s^2 \a_{weak})$
\beg  \plabel{erg}
\D \G_{boundstate} = -4 i \G \< \< \l^- \otimes (\pss-m) \>\>=
-\G_0 \frac{\< \p\,^2 \>}{2 m^2}
\ee
Now it is clear from the preceding argument that it applies not only
to the toponium system but also to all other
systems which can be described as weakly bound fermion- antifermion
system with unstable components. Another example which is known for a long
time is provided by the muonium system where the above result has also been
obtained first by explicit calculation \pcite{Ueber}.  We conclude that the
formalism developed here and especially the use of the identity \pref{WI}
simplifies the problem of the bound state
correction to the decay width in a profound way. It is now possible to
clearly isolate the underlying cancellation mechanism which automatically
gives a gauge independent result which can be interpreted as time dilatation
alone. However, because we may now also include the scalar part of the
self energy function and since $K_0$ can be any kernel, the result
\pref{erg} is found to comprise \pcite{topdec} and \pcite{Ueber} as
special cases of a more general theorem: The leading bound state corrections
for weakly bound systems of unstable fermions (with decays like
$t \to b + W^+, \mu^- \to e^-+\bar{\n}_e+\n_{\mu}$) are {\it always}
of the form \pref{erg}.
 Among the consequences we especially not that \pref{erg}
may be even safely applied to toponium even if one uses e.g.a
renormalization group improved potential.

\vspace{1cm}

{\bf Acknowledgement:} This work has been supported by the Austrian Science
Foundation (FWF), project P10063-PHY within the framework of the EEC- Program
"Human Capital and Mobility", Network "Physics at High Energy Colliders",
contract CHRX-CT93-0357 (DG 12 COMA).

\small
\renewcommand{\baselinestretch}{1.2} 

\hfil \epsfbox{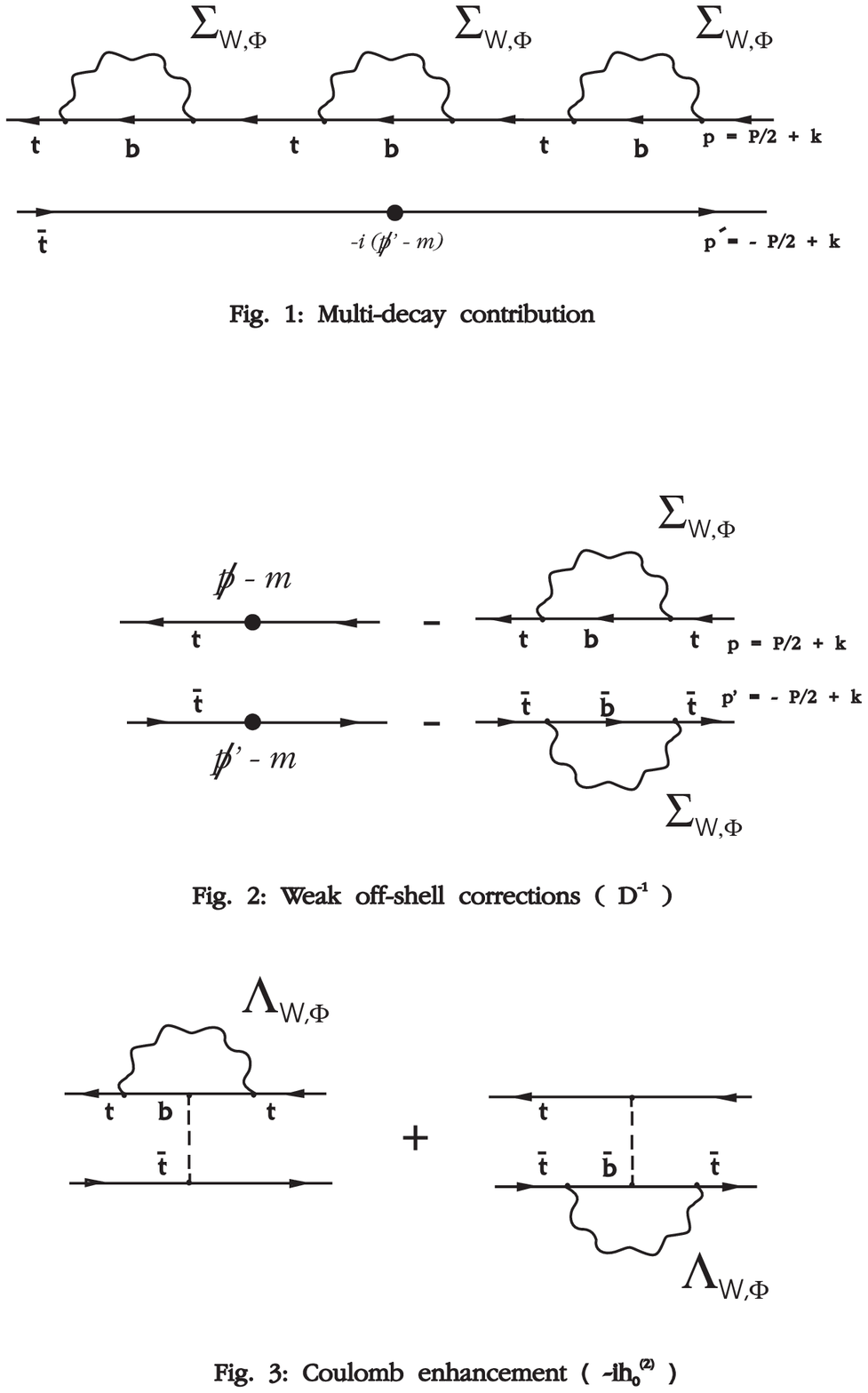}

\end{document}